\documentclass[iop]{emulateapj}  

\usepackage{graphicx}
\usepackage{natbib}
\usepackage{txfonts}
\bibpunct{(}{)}{;}{a}{}{,} 
\shorttitle{Magnetic field in atypical prominences}
\shortauthors{Levens et al.}

\begin{document}
\title{Magnetic  field in atypical  prominence structures: Bubble, tornado and eruption}


\author{P. J. Levens}
	\affil{SUPA School of Physics and Astronomy, University of Glasgow, Glasgow, G12 8QQ, UK}
	\email{p.levens.1@research.gla.ac.uk}

\author{B. Schmieder}
	\affil{LESIA, Observatoire de Paris, PSL Research University, CNRS, Sorbonne Universit\'{e}s, UPMC Univ. Paris 06, Univ. Paris Diderot, Sorbonne Paris Cit\'{e}, 5 place Jules Janssen, F-92195 Meudon, France}
	\email{brigitte.schmieder@obspm.fr}

\author{A. L\'{o}pez Ariste}
	\affil{IRAP - CNRS UMR 5277. 14, Av. E. Belin. 31400 Toulouse. France}
	\affil{Universit\'{e} de Toulouse, UPS-OMP, Institut de Recherche en Astrophysique et Plan\'{e}tologie, Toulouse, France}

\author{N. Labrosse}
	\affil{SUPA School of Physics and Astronomy, University of Glasgow, Glasgow, G12 8QQ, UK}

\author{K. Dalmasse}
	\affil{CISL/HAO, National Center for Atmospheric Research, P.O. Box 3000, Boulder, CO 80307-3000, USA}

\and

\author{B. Gelly}
	\affil{CNRS UMR 3718 THEMIS, La Laguna, Tenerife, Spain}


   \date{Received ...; accepted ...}

 
  \begin{abstract}
Spectropolarimetric observations of prominences have been obtained with the THEMIS telescope during four years of coordinated campaigns. Our aim is now to understand the conditions of the cool plasma and magnetism in `atypical' prominences, namely when the measured inclination of the magnetic field departs, to some extent, from the predominantly horizontal field found in `typical' prominences. What is the role of the magnetic field in these prominence types? Are plasma dynamics more important in these cases than the magnetic support? We focus our study on three types of `atypical' prominences (tornadoes, bubbles and jet-like prominence eruptions) that have all been observed by THEMIS in the He I D$_3$ line, from which the Stokes parameters can be derived. The magnetic field strength, inclination and azimuth in each pixel are obtained by using the Principal Component Analysis inversion method on a model of single scattering in the presence of the Hanle effect. The magnetic field in tornadoes is found to be more or less horizontal, whereas for the eruptive prominence it is mostly vertical. We estimate a tendency towards higher values of magnetic field strength inside the bubbles than outside in the surrounding prominence. In all of the models in our database, only one magnetic field orientation is considered for each pixel. While sufficient for most of the main prominence body, this assumption appears to be oversimplified in atypical prominence structures. We should consider these observations as the result of superposition of multiple magnetic fields, possibly even with a turbulent field component.
\end{abstract}


   \maketitle
%

\section{Introduction}


With the current armada of spacecraft observing the Sun we  have an accurate view of solar phenomena occurring in the corona. 
Movies obtained with high spatial and temporal resolution by imagers such as the Solar Optical Telescope \citep[SOT;][]{Suematsu2008,Tsuneta2008} on the \textit{Hinode} satellite \citep{Kosugi2007} and the Atmospheric Imaging Assembly on the \textit{Solar Dynamics Observatory} \citep[AIA, \textit{SDO};][]{Lemen12} allowed us to discover the incredible dynamic nature of prominences, even the quiescent ones:  apparent up and 
down flows in quasi-vertical structures, rising bubbles and apparently rotating - tornado-like - structures \citep{Dudik2012,Orozco2012,Wedemeyer2013,Berger2014,Su2014}. 
{These observational characteristics lead us to the question: are the dynamics within the prominence more important than the magnetic field for prominence formation and stability?} {This would require a plasma-$\beta > 1$}.
Spectroscopy is useful to analyse  the real plasma motion and its physical conditions \citep{2010SSRv..151..243L}.  
 Rotation around a central axis has been detected in a tornado-like structure, in hot plasma ($\log{T} > 6$) surrounding the prominence legs by 
 the Extreme-ultraviolet Imaging Spectrometer \citep[EIS;][]{Culhane07}, a subsystem of the \textit{Hinode} satellite \citep{Su2014,Levens2015}. 
Using H$\alpha$ spectra obtained with the Multi-channel Subtractive Double Pass (MSDP) instrument at the Meudon Solar Tower {it has been seen that a high level of dynamics in a prominence can derive from magnetic fine structures that are weakly magnetised \citep{Gunar2012}.}
However, it has been shown that the velocity vectors are not aligned with the apparent vertical structures in prominences, as the movies from \textit{SDO}/AIA and \textit{Hinode}/SOT suggest, but in fact have a significant 
angle with  respect to the vertical \citep{Schmieder2010}.  In hedgerow prominences, \citet{Chae10} suggested that the 
  descending observed  knots are supported by horizontal magnetic fields against gravity, even when they are moving downwards, and the complex variations of their descent speeds should be attributed to small imbalances between gravity and the force of magnetic tension.
The IRIS {satellite} \citep{DePontieu2014} provides tremendous data on prominences. It has observed a number of quiescent and eruptive prominences in the chromospheric \ion{Mg}{2} h and k lines \citep{Schmieder2014,Liu2015,Vial2016,Levens2016}.
The highly dynamic plasma observed, even in quiescent prominences, could answer the question of short scale height of the plasma pressure 
compared with the common height of prominences \citep{Schmieder2014}. 

Theoretical models of prominences are mainly based on static
structures  \citep{Aulanier1998,vanBallegooijen2004,Dudik2008,Mackay2010}. The plasma would be sustained in a pile of magnetic dips in sheared arcades, or in
twisted flux ropes due to the magnetic tension force or the presence of a tangled magnetic field on small scales \citep{Lopez2006,Lites2010,vanBallegooijen2010}. 
Only recently has it been possible to create  MHD models with a self-consistent plasma-carrying  flux rope and producing in situ condensation forming a prominence \citep{Xia2014,Terradas2015}.

All of these models are consistent with
previous measurements of the magnetic field in prominences. Polarimetry of prominences was achieved in the 1980s \citep{Leroy1984} which showed that the 
prominence magnetic field vector was almost horizontal (60$^\circ$ to 90$^\circ$ from the vertical). The {diagnostic techniques}  developed during this time period involved 
only the Hanle effect, neglecting the Zeeman effect \citep{Bommier1994}. It now appears that the second effect cannot be 
ignored. New inversion codes have been  developed to  include both of these effects, and they are now being applied to new sets of 
full Stokes  vector observations \citep{LAC02,Casini2003,Lopez2007,Lites2014}.  
Since 2012, the French  T\'elescope H\'eliographique pour l'Etude du Magn\'etisme et des Instabilit\'es Solaires (THEMIS)  in the Canary Islands  with the MulTi Raies (MTR) mode has been observing prominences during a number of international campaigns. More than 200 observations of prominences have been made in the \ion{He}{1} D$_3$ line, from which  statistics have been presented -- during the IAU S305 symposium \citep{Lopez2015,Schmieder2015} -- and case studies have been published \citep{Schmieder2013,Schmieder2014}.  The principal result is {still} that the magnetic field is mainly horizontal in prominences.
{However, not} all structures that are observed over  the solar limb are  prominences in the classical sense. To identify a prominence  it is necessary to have observations of the corresponding H$\alpha$  filament  overlaying a magnetic inversion line a few days before its passage across the limb. 

{In this paper we present and discuss magnetic field maps of atypical prominences that have been observed with the THEMIS/MTR instrument during these campaigns. These prominences are not typical prominences, in the sense that the measured magnetic field inclination presents some departure from the predominantly horizontal field found in typical prominences. The examples presented here were identified as quasi-vertical tornadoes or jet-like prominences. } We also measure the magnetic field in bubbles below prominences, which is a very important, yet still unknown, parameter in the simulation of Rayleigh-Taylor buoyancy instabilities, which are responsible for the formation of bubbles \citep{Ryutova2010,Berger2014}.

In Section 2 we describe the polarimeter  operating in the  THEMIS telescope in the Canary Islands,{ as well as the data reduction and analysis,} in Section 3 we discuss the atypical prominences with 
  context images obtained by both ground-based and space-based instruments (i.e. \textit{SDO}/AIA filters and \textit{Hinode}/EIS  spectrograph and \textit{Hinode}/SOT). We present examples of each particular case (prominences with bubbles, tornadoes and  an eruptive prominence) and discuss the role of magnetic field in the frame of formation and stability of prominences.


\section{THEMIS}

\subsection{Instrument}

\begin{figure}
   \begin{center}
    \includegraphics[width=0.3\textwidth,clip=]{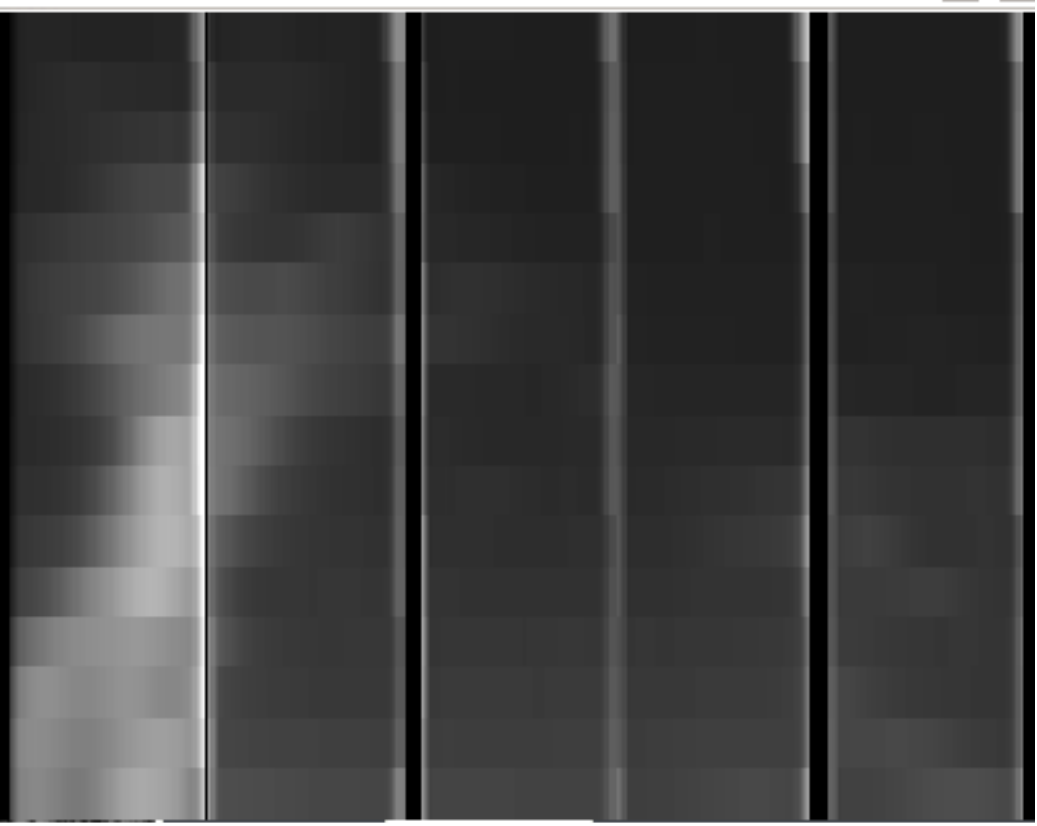}
    \includegraphics[width=0.31\textwidth,clip=true, trim=4cm 3cm 4cm 3cm]{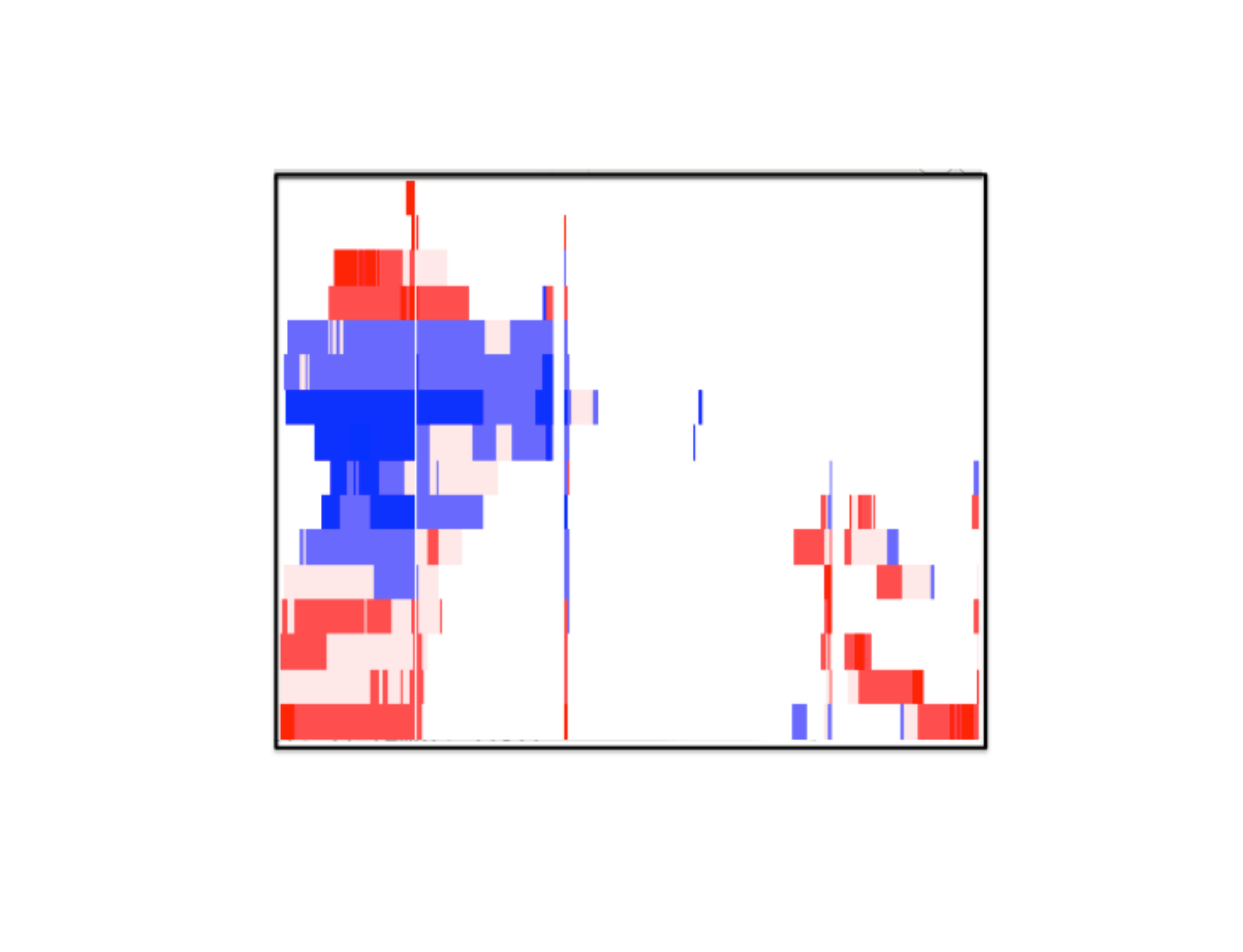}
\caption{Observation from May 23, 2014  at 15:24 UT  obtained  by THEMIS in the \ion{He}{1} D$_3$ line. \textit{Top:} Intensity map in \ion{He}{1} D$_3$. \textit{Bottom:}  Doppler shift map. There is a velocity difference of $\sim$2 km s$^{-1}$ between the red and blue. {The pixel size is 0.23\arcsec\ in $x$ and  2\arcsec\ in $y$.}   }
         \label{themis_D3}
	\end{center}
   \end{figure}

The THEMIS/MTR { instrument }allows us to make spectropolarimetric measurements in the \ion{He}{1} D$_3$ line in prominences \citep{LARS00,Paletou2001}. The spectrograph slit is commonly 
orientated parallel to the limb. 
The double-beam polarimetry requires the use of a grid mask {presenting} three segments 15.5\arcsec\ wide along the slit, 
but {which masks} regions of 17\arcsec\ between each segment. The masked regions allow us to obtain a double image with opposite polarization {to the unmasked ones}, but {force the observer}
 to scan in the direction along the slit by one step of 15\arcsec\ to {fill the holes} in order to get continuous coverage  of the prominence. This mode  of  scanning  along the
 slit  introduces jumps and dark lines to the intensity images and Dopplershift maps made using the \ion{He}{1} D$_3$ line, such as those presented as an example in Figure \ref{themis_D3}. The zero velocity in the lower panel of Figure \ref{themis_D3} is arbitrary, as it is calculated by taking the average across the whole raster. Therefore it does not take into account any potential movement of the whole prominence, and should be viewed as a relative Doppler shift. There is a gradient of velocities from the bottom to the top of the tornado and an increase of 1.8 km s$^{-1}$.

In addition to that scan along the slit, a {more traditional} scan perpendicular to the slit with steps of 2\arcsec\ from the limb to the top of the prominence was performed. Fields 
of view of 120\arcsec\ $\times$  20--40\arcsec\ are covered in about half an  hour  to an hour with 
an  exposure time of 2 seconds per Stokes parameter and scan position. Full polarimetry with beam-exchange is done with a modulation cycle of 6 images, 
spanning the three Stokes parameters with either positive or negative
sign measured in every beam, and the simultaneous double beam measuring the opposite sign. Each Stokes parameter is thus measured in the same camera 
pixel at 
two different times and in two different pixels at the same time. This symmetry of measurements results in  high  polarimetry precision  and a reduction of the 
systematic errors  to a fourth order perturbation 
of the signal. Each cycle was repeated five times to increase signal-to-noise (S/N) ratios.


\begin{figure*}
\begin{center}
\includegraphics[width=0.85\hsize]{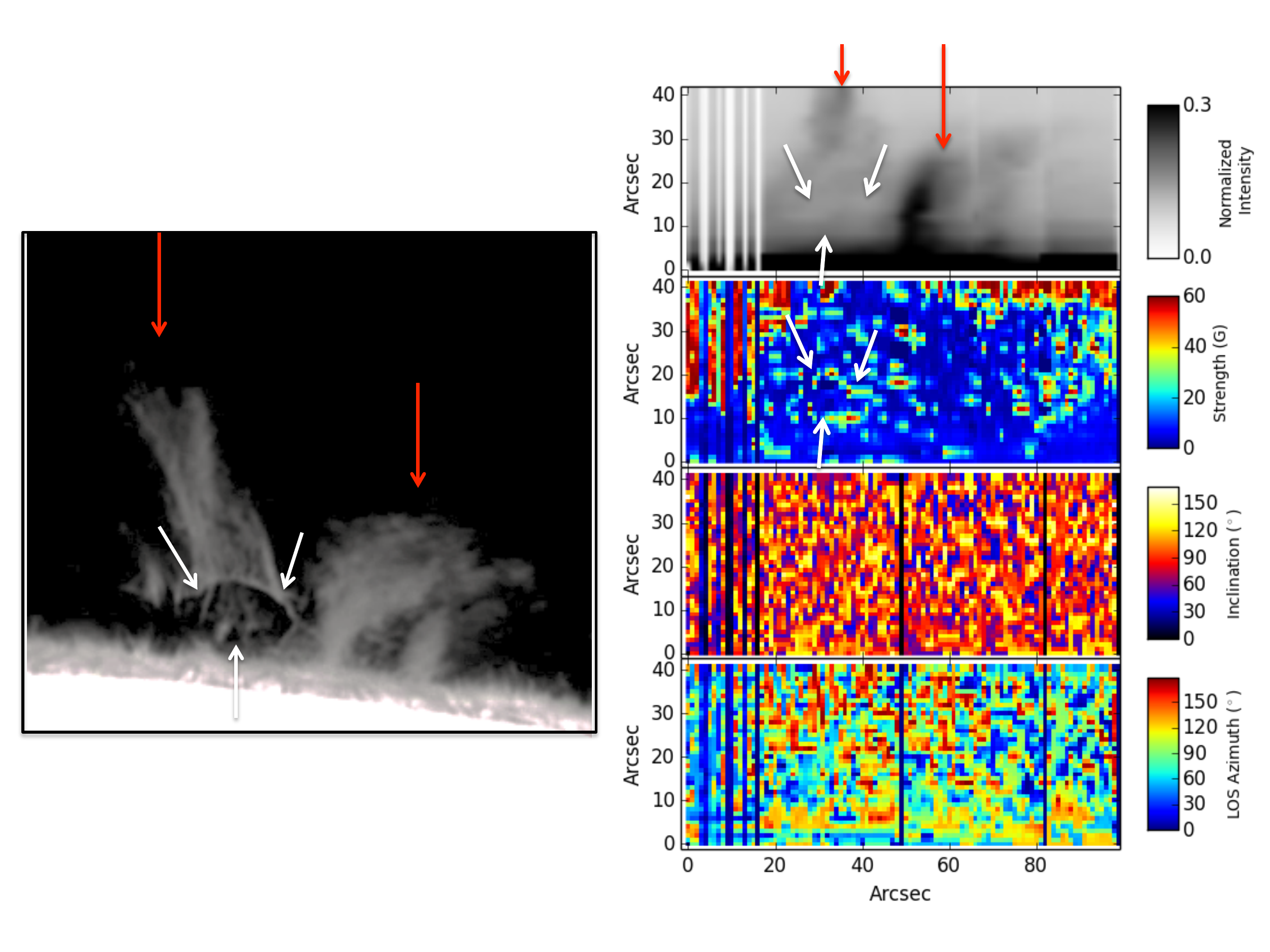}
\caption{Bubble observed on the 24 May, 2014. \textit{Left:} H$\alpha$ image from the NSVT instrument (movie available online, image courtesy of Xu Zhi). The black arrow indicates the location of the bubble. \textit{Right:} THEMIS/MTR maps made using the \ion{He}{1} D$_3$ line. From top to bottom: Intensity (reversed grey scale), magnetic field strength, inclination, and azimuth. White arrows indicate {strong magnetic fields around the bubble}. {The white arrows indicate the location of the bubble in the NSVT and in the THEMIS images. The red arrows are pointing the two different structures of the prominences.}}
\label{fig2new}
\end{center}
\end{figure*}

\begin{figure*}
\begin{center}
\includegraphics[width=0.8\hsize]{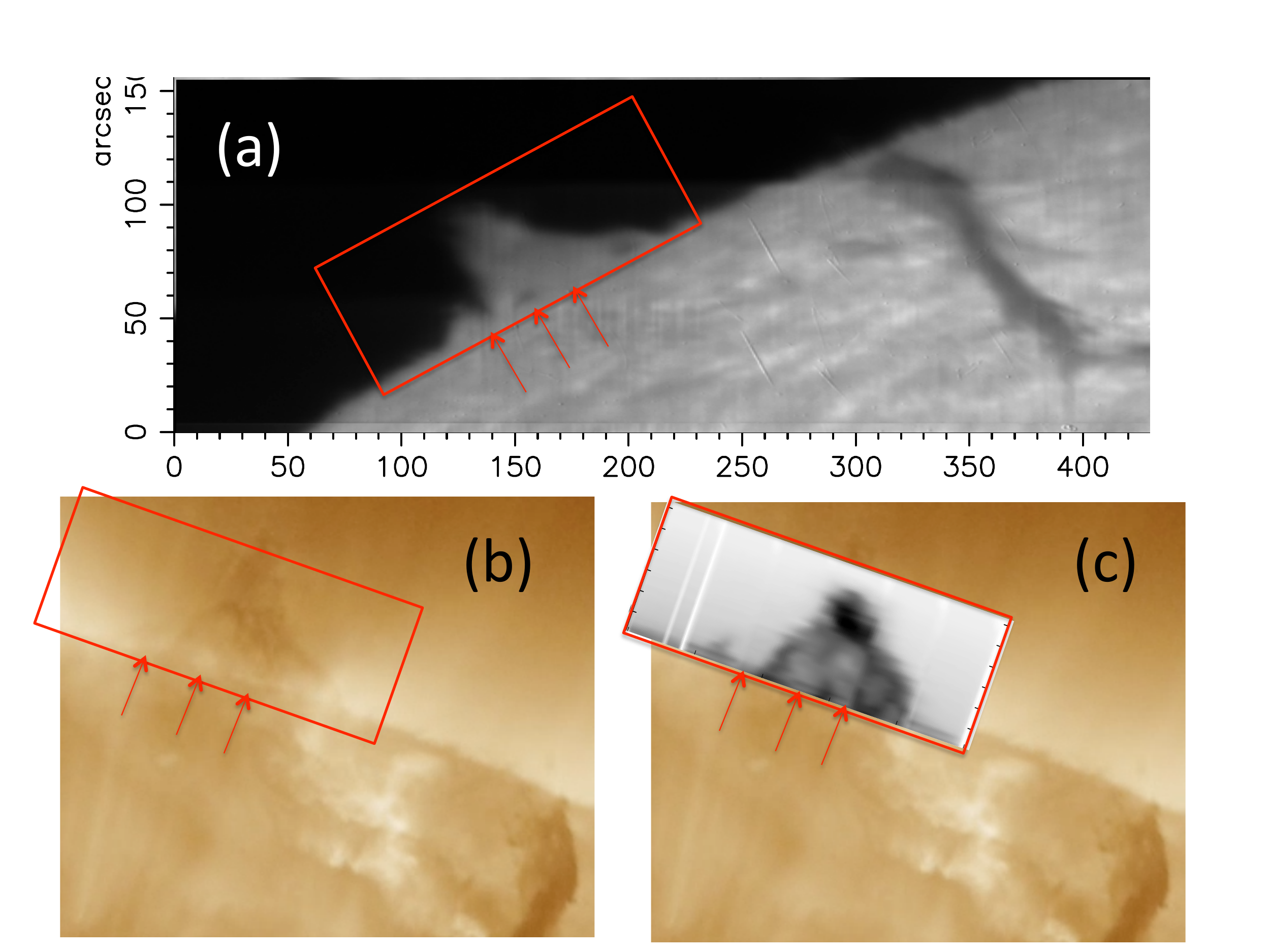}
\hspace{-0.5cm}
\includegraphics[width=0.8\hsize, clip=true, trim=0 4cm 0 2cm]{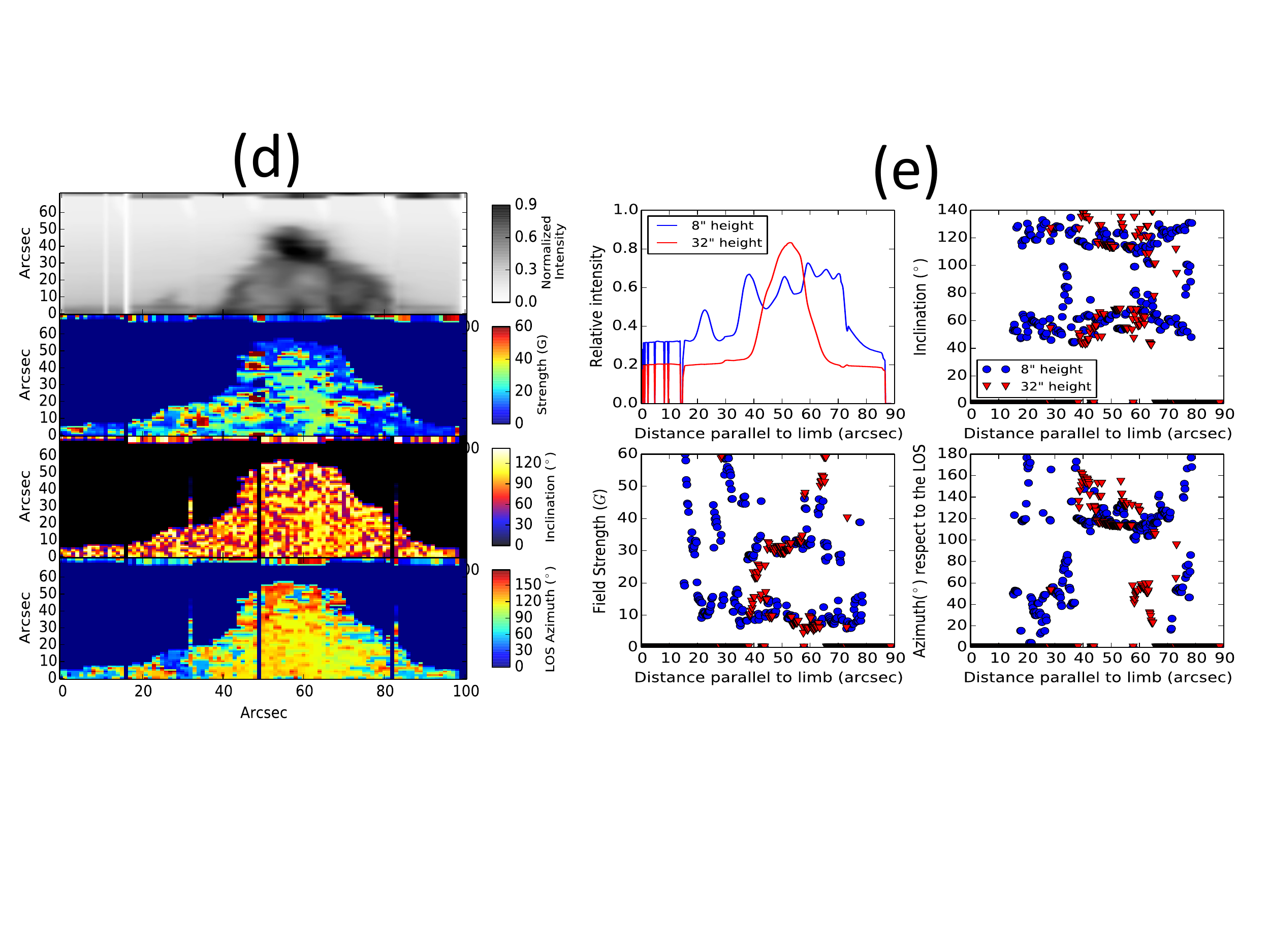}
\caption{Bubbles under a prominence observed on the 7 May, 2015.{\textit{(a)} H$\alpha$ image from the MSDP instrument at Meudon. \textit{(b)} and \textit{(c)} AIA 193 \AA\ image of the prominence.  Red arrows indicate the {approximate} locations of the bubbles. Panel \textit{(c)} has the \ion{He}{1} D$_3$ intensity image superimposed onto AIA 193 \AA. \textit{(d)} THEMIS maps for the prominence and bubbles using the \ion{He}{1} D$_3$ line. From top to bottom: Intensity (reversed grey scale)}, magnetic field strength, inclination, and azimuth. \textit{(e)} Cuts through the prominence and bubbles shown in \textit{(c)}; \textit{Upper:} Intensity (left) and inclination (right). \textit{Lower:} Field strength (left) and azimuth (right).}
\label{fig3new}
\end{center}
\end{figure*}

\begin{figure*}[t!]
\begin{center}
\includegraphics[width=0.85\hsize]{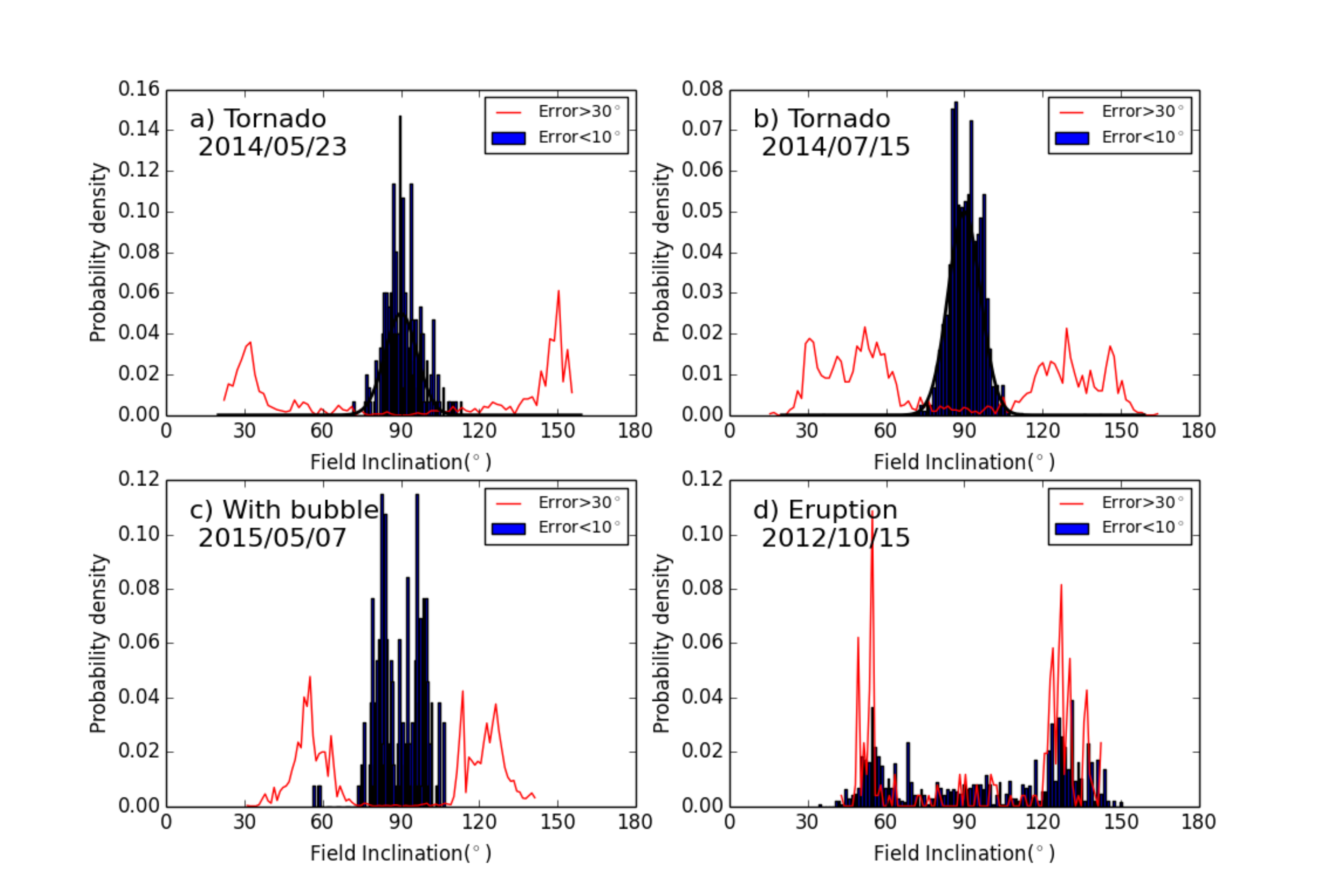}
\caption{Histograms of magnetic field inclination with respect to the local vertical for the different prominence types presented here. \textit{(a)} Tornado observed on 23 May, 2014. \textit{(b)}  Tornado observed on 15 July 2014. \textit{(c)} Prominence with bubbles observed 7 May, 2015. \textit{(d)} Eruption, observed on 15 October, 2012. Blue bars show points where the error in the inclination is less than 10$^\circ$, the red lines indicate where the error is larger than 30$^\circ$. In \textit{(a)} and \textit{(b)} a Gaussian of FWHM $10^{\circ}$ has been drawn representing the expected error distribution from the inversion.}
\label{fig4new}
\end{center}
\end{figure*}

\begin{figure*}[t]
\begin{center}
\includegraphics[width=0.30\hsize,clip=true,trim=0 5cm 2cm 5cm]{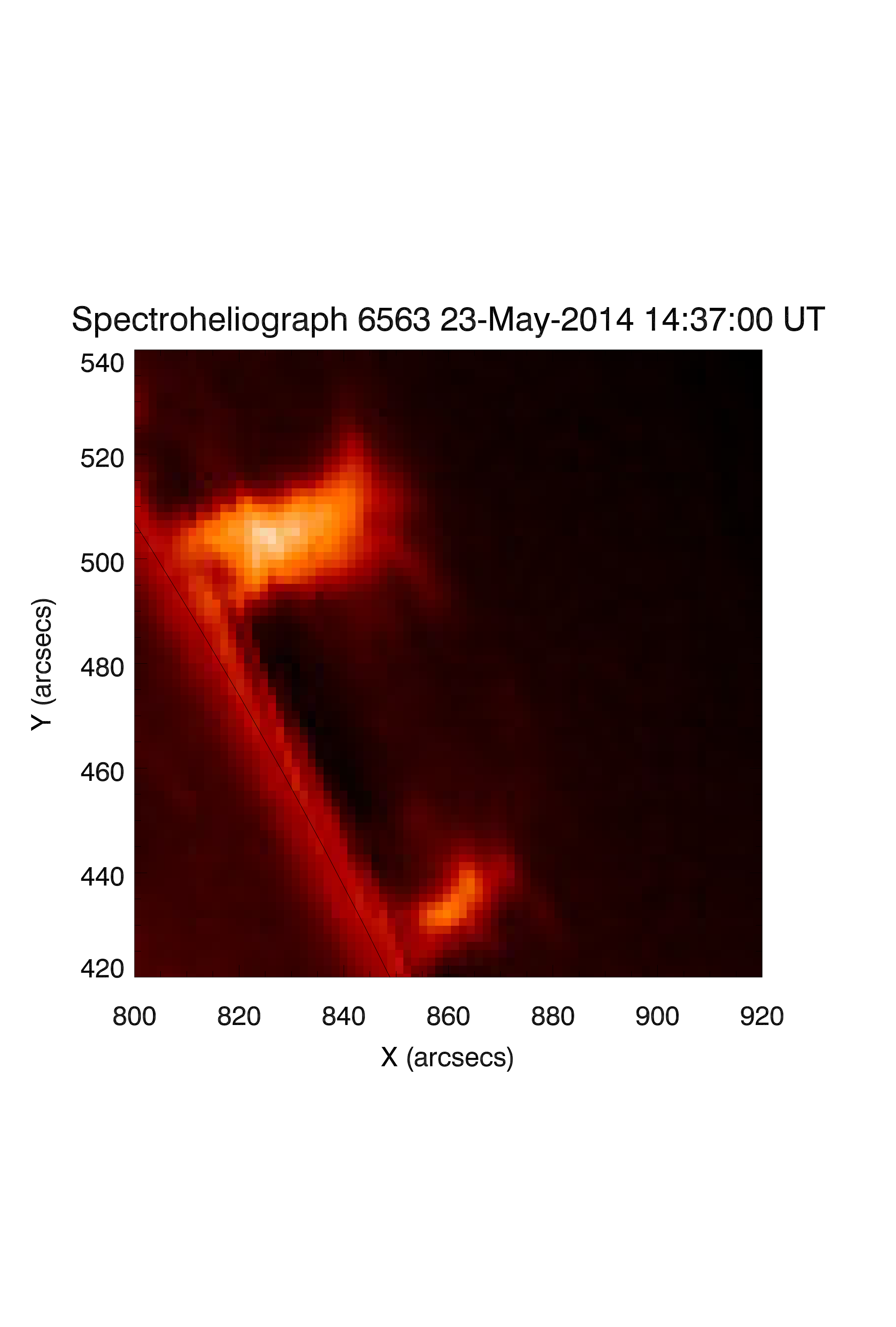}
\includegraphics[width=0.30\hsize,clip=true,trim=0 5cm 2cm 5cm]{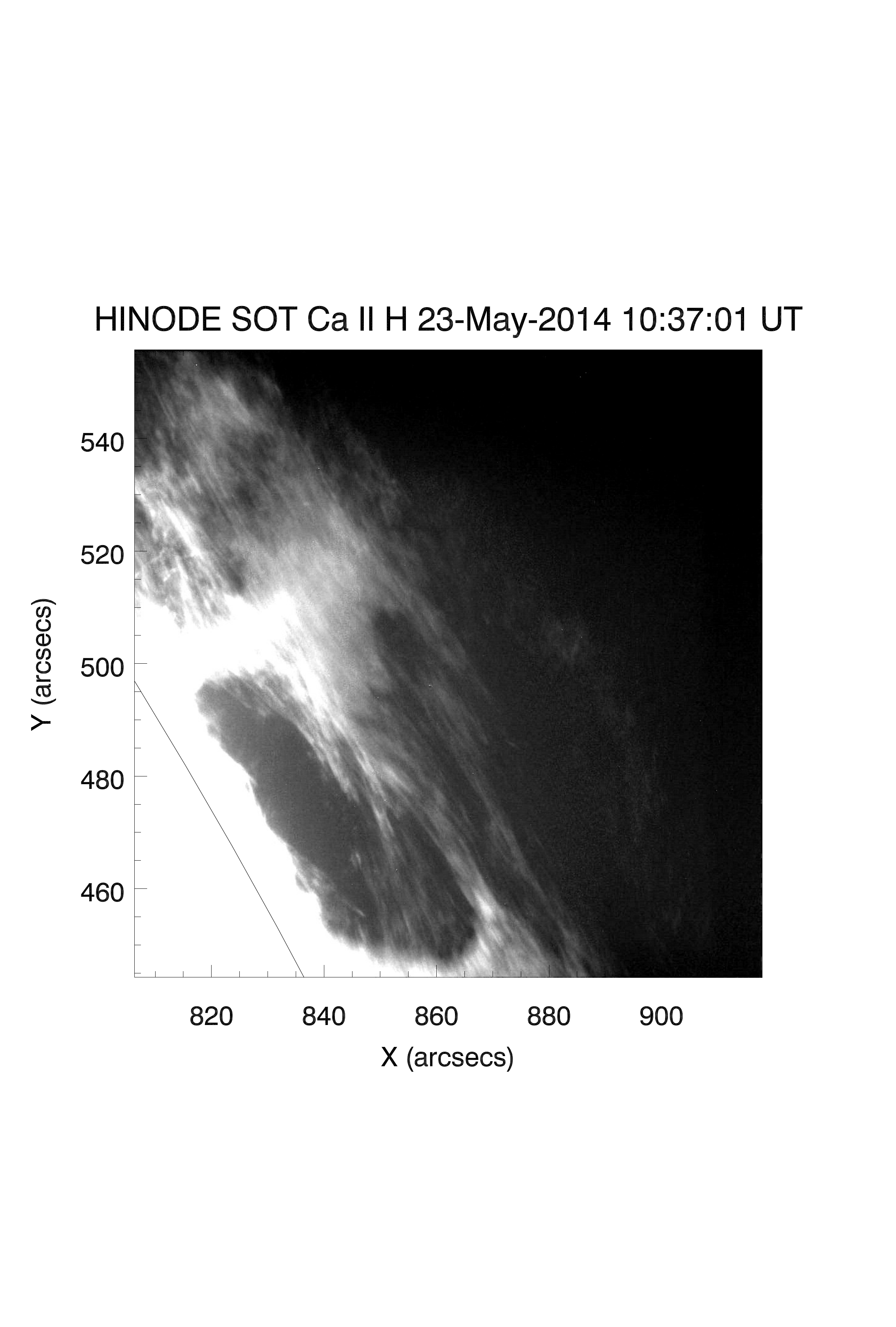}
\includegraphics[width=0.25\hsize,clip=true,trim=0 2.5cm 2cm 6cm]{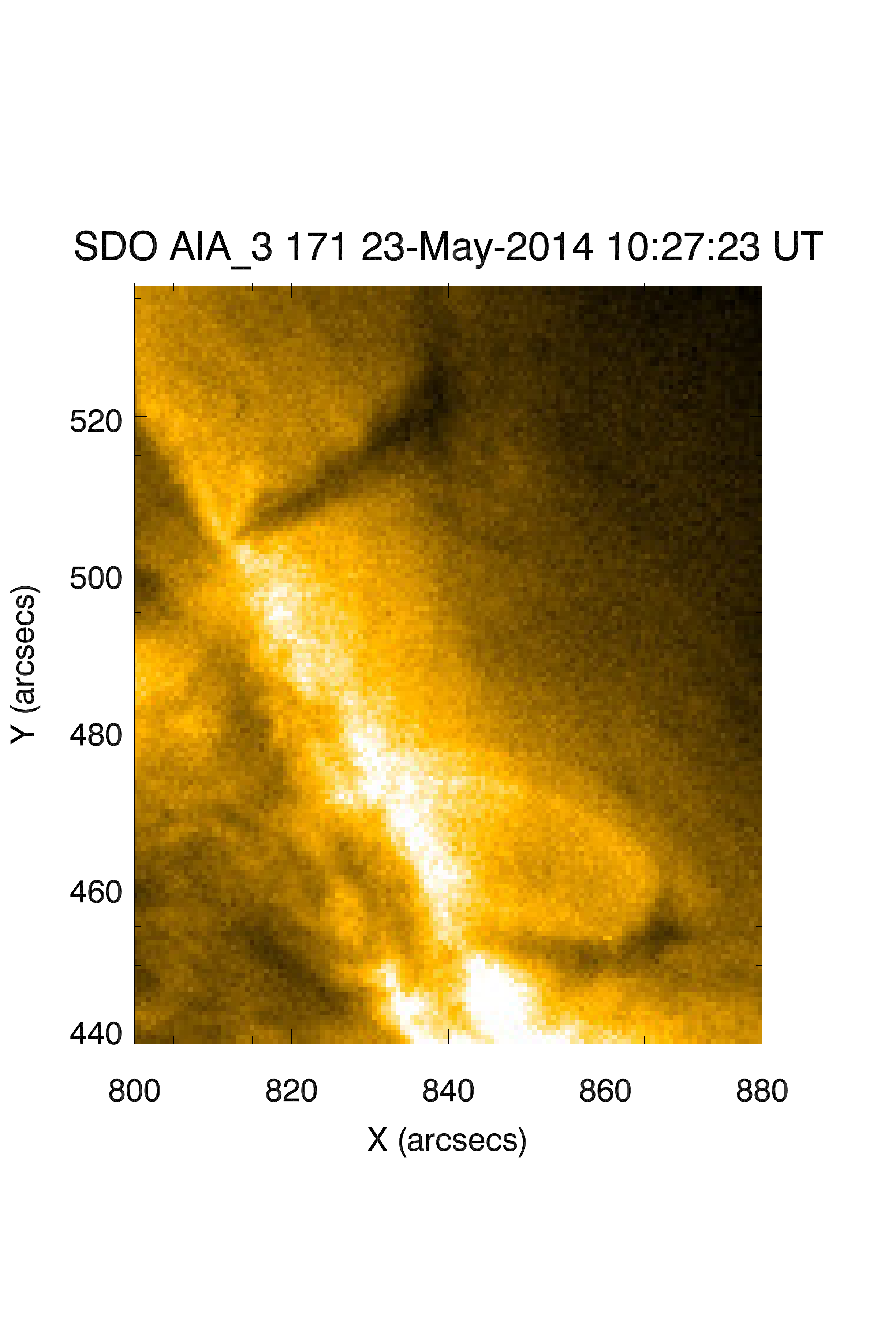}
\includegraphics[width=0.30\hsize,clip=true,trim=0 0 2cm 5cm]{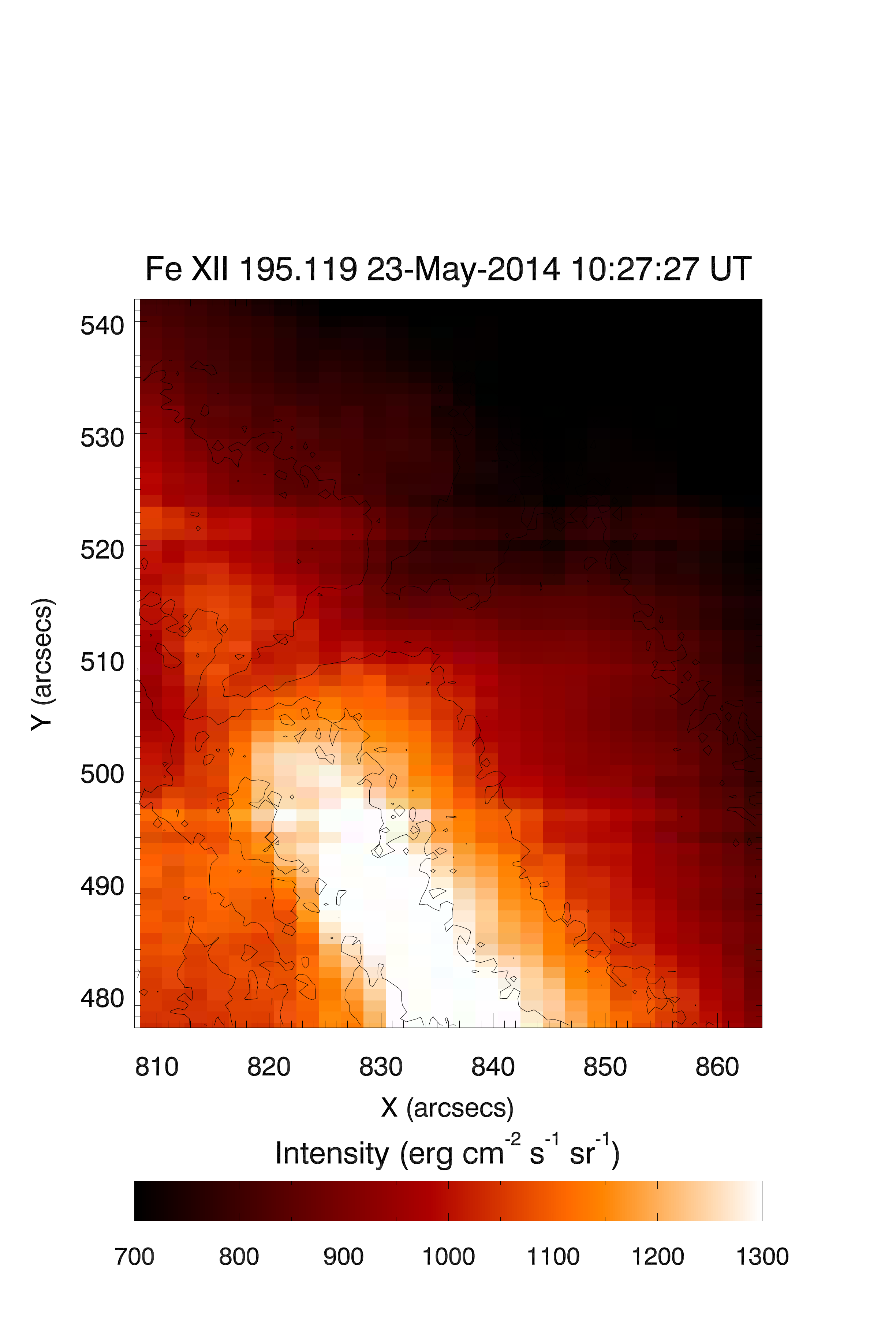}
\includegraphics[width=0.30\hsize,clip=true,trim=0 0 2cm 5cm]{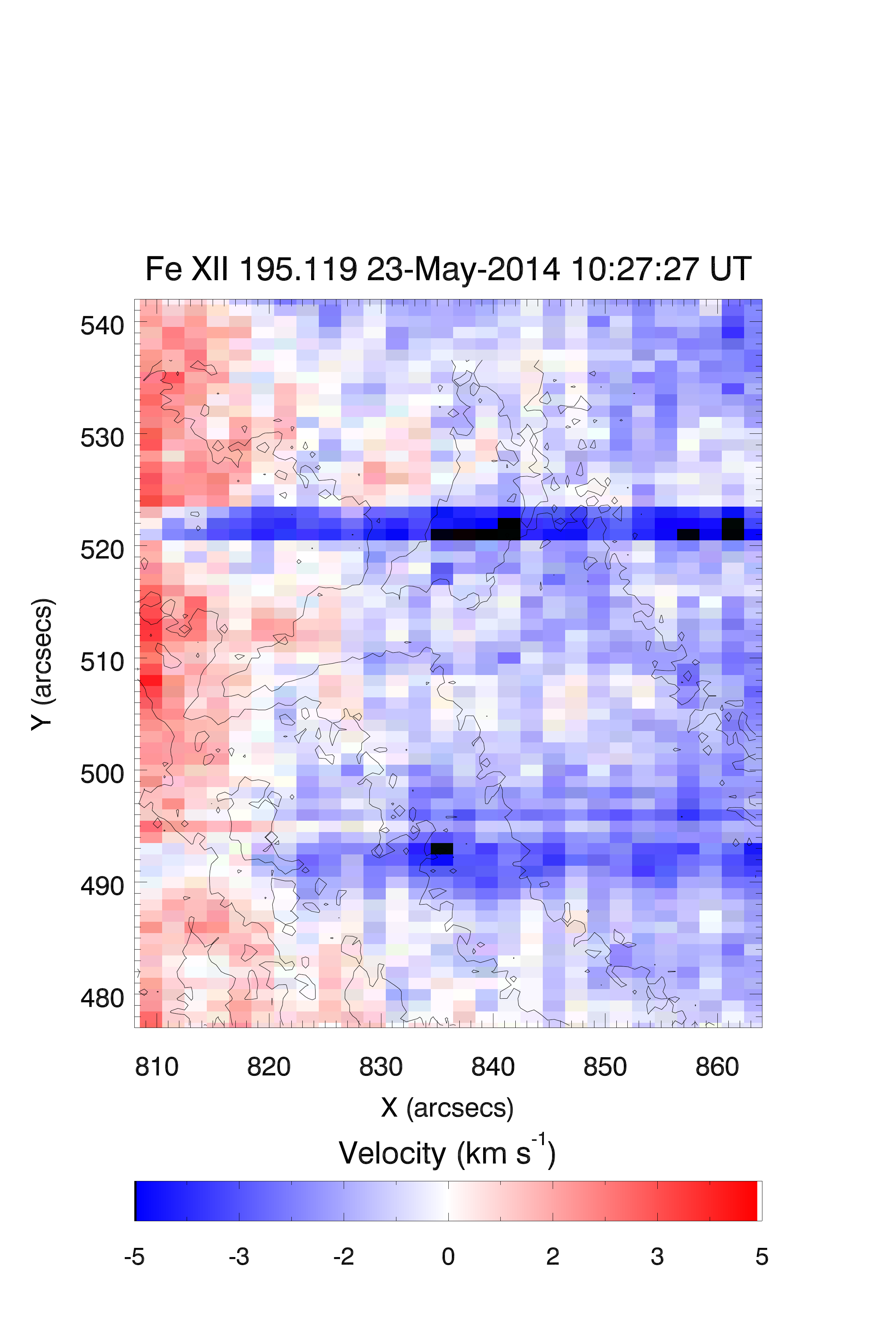}
\caption{Tornado observed on 2014 May 23. \textit{Top:} Intensity maps in H$\alpha$ from the Meudon Solar Survey instrument (left, tornado is northern column), \ion{Ca}{2} from \textit{Hinode}/SOT (centre), and AIA 171 \AA\ image (right, tornado is northern column). A movie of the AIA 171\AA\ image is available online. \textit{Bottom:} Intensity (left) and velocity (right) maps made using the \ion{Fe}{12} 195.119 \AA\ line observed by \textit{Hinode}/EIS. These maps show a zoomed-in region of interest around the tornado. The horizontal stripes in the EIS images are due to warm pixels on the EIS CCD. Contours are taken from AIA 171 \AA\ image shown. Note: all images have different spatial scales.}
\label{EIS}
\end{center}
\end{figure*}

\subsection{THEMIS spectropolarimetry}

Raw data from the THEMIS/MTR observations has been reduced using the DeepStokes procedure, which is outlined in \citet{LAARMSDG09}. Data reduction included flat-fielding, dark current and
 bias subtraction, wavelength 
calibration and a careful handling of the polarization signals. The results of the data reduction are cubes of spectra for  the \ion{He}{1} D$_3$ line in
intensity,
linear polarization (both $Q$ and $U$) and circular polarization ($V$), for all points along the slit and all positions of the double scan. S/N ratios are higher than
$10^3$ at the \ion{He}{1} D$_3$ line core in central parts of the prominence for each of the Stokes parameters.

The Stokes profiles are taken as the inputs of an inversion code  based on Principal Component Analysis  (PCA) \citep{LAC02,Casini2003}. This code efficiently compares the observed
profiles to
those in a database, which was generated {using synthetic} models of the polarization profiles of  \ion{He}{1} D$_3$ (see \citet{LAC02} for more details). The  comparison 
is made independently pixel by pixel. The database  we used contains around 90000 profiles, computed as the emission of 
a single He atom in its triplet state, modelled with the 5 levels of lower energy of the He triplet system. The atom is polarized by the anisotropic radiation {from the disc} below the prominence. The atomic polarisation depends on the height of the {scattering volume, which} is one of the free parameters of the model. Collisions are not taken into account. The atomic
polarization of the helium atom is modified by a single vector magnetic field with free strength, inclination and azimuth. The Hamiltonian of the atom includes all 
terms with its Zeeman sublevels splitting linearly with the magnetic field. We solve the density matrix  of the atom in statistical equilibrium; the solution 
contains all populations and quantum coherences, including 
atomic alignment and orientation, for all levels involved in the He triplet atom model. The Hanle effect of every level is thus computed, as well as the
Zeeman effect. From the resulting populations and coherences we compute the polarization 
dependent emission terms,  in whatever direction we are observing. The scattering angle is thus a free parameter of the model too. Several million
profiles are  computed and  used to build  the database, keeping just those which are different enough  and rejecting others so that the database fills the space of 
possible profiles as homogeneously as possible, while keeping  its size  small.

The \ion{He}{1} D$_3$ line at 5876 \AA\ is {seen as} a doublet, where the sensitivity of each component to the Hanle effect is different. Saturation of these lines is not an issue under prominence conditions.
These facts allow us to better constrain ambiguities and error bars. This is in contrast with the other important helium line in prominences,  as the \ion{He}{1} 10830 \AA\ line \citep{Orozco2014}. The analysis 
of the information contained within the two helium lines made by \cite{LAC05} and \cite{Casini2009}, using PCA-based inversion codes, demonstrated 
that the \ion{He}{1} D$_3$ line better constrains the 
solution. {In particular,} this results in better determined field strengths and inclinations that seldom suffer from the 90$^\circ$ ambiguity.
Furthermore, the PCA  inversion code, {thanks to its global search,}  provides a solution and  error bars for every pixel in the observation. This
error bar is computed as the standard deviation over all solutions in the database that fit the observation within predefined margins. They include noise-related 
errors, but they also point to ambiguities. Well inverted profiles can be declared free of ambiguities other than the ubiquitous 180$^\circ$ ambiguity in the 
azimuth of the field (in the reference frame of the observer). Because of this, the few situations in which the 90$^\circ$ ambiguity could appear in our solutions, would be conspicuous due to large error bars of 90$^\circ$ or more. We have not found problems of this kind in {the datasets presented in this work}, indicating that {they are} free from the 90$^\circ$ ambiguity.

After comparison of an observed profile with all those in the database, the most similar profile is kept as the solution.  The main output is the vector magnetic field, and we check that the {other} free parameters of the model  
(height above the photosphere and scattering angle) are in a typical range where the polarization of \ion{He}{1} D$_3$ is not so dependent on their values.
The retrieved heights are in the ballpark of
the observed heights, but they do not carry any new information. As said above, height and scattering angle are free parameters of the model used for inversion. However, the simulations of  \citet{LAC02}  show that the uncertainties in those parameters are quite large, {even if} the other magnetic parameters are still determined reliably. 

Error bars are determined for the magnetic parameters by doing
 statistics on all other models that are sufficiently similar to the observed ones, but not as similar as the one that is selected as the solution. It is important to
stress that, although there is always one case in the database that is the most similar to the observed one, this does not mean that it is a good fit. It is therefore important to keep a measure of how similar they are, and to check that all conclusions about the magnetic field strength or 
orientation are based upon sets of profiles that { correctly fit the observations.}




\section{Bubbles below prominences}

Bubbles underneath  prominences are frequently observed by \textit{Hinode}/SOT and SDO/AIA  \citep{Berger2008,Dudik2012,Gunar2014}.  {They appear as dark structures in H$\alpha$ and bright in AIA 171 \AA\ and 193 \AA. } {The dynamic} behaviour of  prominences was studied by \citet{Berger2010,Berger2011}, who described the evolution of bubbles  producing small-scale plumes rising upwards. These plumes could transport hot plasma upwards, which could then condense and fill the cavity.
The magnetic topology of prominence bubbles was studied by \citet{Dudik2012} and the possible mechanism for triggering the dynamic behaviour of bubbles and plumes was modelled by \citet{Hillier2011}.
\citet{Gunar2014} argued that the  prominence bubbles could be formed due to perturbations in the magnetic field by parasitic bipoles on the solar surface, causing them to be devoid of magnetic dips.
{The bubbles would be surrounded by a separatrix, a thin layer of current where magnetic field can reconnect. The rise of the dome-like bubble would be due to the magnetic pressure being larger  {in and around}  the bubble than in the {prominence itself}, and the plasma moving up and down would be caused by reconnection along the separatrix.}

On May 24 2014, a bubble was identified underneath a prominence by the New Vacuum Solar Telescope (NVST) in Kunming, as seen in Figure \ref{fig2new} (\textit{left panel}) \citep{Shen2015}. {THEMIS observed this object at around the same time, and}
Figure \ref{fig2new}  (\textit{right panel}) presents the maps obtained after inversion  of the Stokes parameters recorded in the \ion{He}{1}  D$_3$  line: intensity in reverse colour, magnetic field strength, inclination, azimuth. The origin for the inclination is the local vertical, and the origin of the azimuth is the line of sight (LOS), in a plane containing the LOS  and the local
vertical.  {Unfortunately, observation conditions during this sequence were far from ideal and
the S/N ratio is low all over the prominence, }in the bubble in particular. It is difficult to detect any signal. However we note a higher magnetic field strength at the edge of the bubble reaching 50 Gauss, shown by white arrows in Figure \ref{fig2new}, a tendency which re-appears in subsequent data for which we lack of the imaging counterpart of this example. 
{This ring of higher magnetic field around the bubble could correspond to a magnetic separatrix \citep{Gunar2014}.}

The second example {of a bubble} was observed on May 7, 2015  {with AIA in 193 \AA\ and in H$\alpha$ (Meudon Solar Tower/MSDP)} (Figures \ref{fig3new} \textit{a}, \textit{b}, \textit{c}). The inclination, field strength and azimuth maps, derived from THEMIS data, are also noisy, as can be seen in the images (Figures \ref{fig3new} \textit{d}). {This time the S/N ratio is not to blame, but rather our modelling of the prominence magnetic field.}
{To enlighten us, we present cuts  parallel to the limb made close to the solar surface, near the base of the prominence (8\arcsec above the surface), together  with another one  higher (32\arcsec) crossing the bubbles at} $x$ = 30\arcsec, 45\arcsec, 55\arcsec\ (Figure \ref{fig3new} \textit{e}). 
 These locations correspond {approximately} to the arrows shown in the AIA  and H$\alpha$ images (Figure \ref{fig3new} \textit{a}, \textit{b}, \textit{c}). {The bubbles move relatively quickly and, due to the fact that the THEMIS image takes one hour to be made, they are not quite at the same places in the AIA image and in the THEMIS image}. 
{Even though this co-alignment is approximate, it suggests that the magnetic field is relatively large in the areas of the bubbles (30 -- 60 G in some pixels at the edges), confirming the possible presence of a separatrix around the bubble, and  that magnetic pressure could be the dominant parameter to explain the rise of  bubbles \citep{Gunar2014}.} 
The inclination shows two distinct behaviours in this prominence that explain the noisy aspect of the images. In Figure \ref{fig4new} \textit{(c)}, a histogram over the whole prominence shows that the correctly fitted profiles (blue bars with errors smaller than 10$^\circ$) concentrate around horizontal values (90$^\circ$). However, contrary to other well studied cases, rather than
a Gaussian peak centered at 90$^\circ$, we observe an unusual bimodal distribution with peaks at roughly 80$^\circ$ and 100$^\circ$, indicating a perturbed magnetic field departing from the stable horizontal geometry. Looking at the bubbles themselves, the inclination shows two distinct values at around 60$^\circ$ and 120$^\circ$. Neighbouring pixels randomly show one or other of these values. As seen in the histogram, these characteristic values are found in solutions with large error bars, larger than 30$^\circ$, which is a symptom of bad inversions. In these cases, the inversion algorithm cannot find a correct solution using a single vector magnetic field and provides an alternative solution which, although incorrect, is systematically the same. As described by \citet{Schmieder2015} and \citet{Lopez2015} such solutions have been associated with turbulent fields, mixed with a dominant, horizontal background field. This is the interpretation that we favour here for the fields found at the bubble pixels.

Given the observational geometry of a bubble, it is clear that most of the light may not be coming from inside the bubble, but rather from the plasma in the walls in front and behind the bubble. What is clear from our data is that, whether inside, in front of or behind, the magnetic field shows a turbulent component on top of the stable horizontal geometry with a tendency to strong fields. If bubbles are understood as originating in parasitic magnetic polarities protruding and disturbing the background magnetic geometry of the prominence, it is not surprising that the full body, even outside the bubbles, presents an unusual magnetic topology. This is perhaps the reason for the slight (barely 10$^\circ$), though ordered, departure from the horizontality that is found 
elsewhere in this prominence. 

\section{Tornado magnetic field}
\label{sec:tornado_B}

 \begin{figure}
   \begin{center}
   \includegraphics[width=\hsize,clip=true,trim=2cm 0 2cm 0]{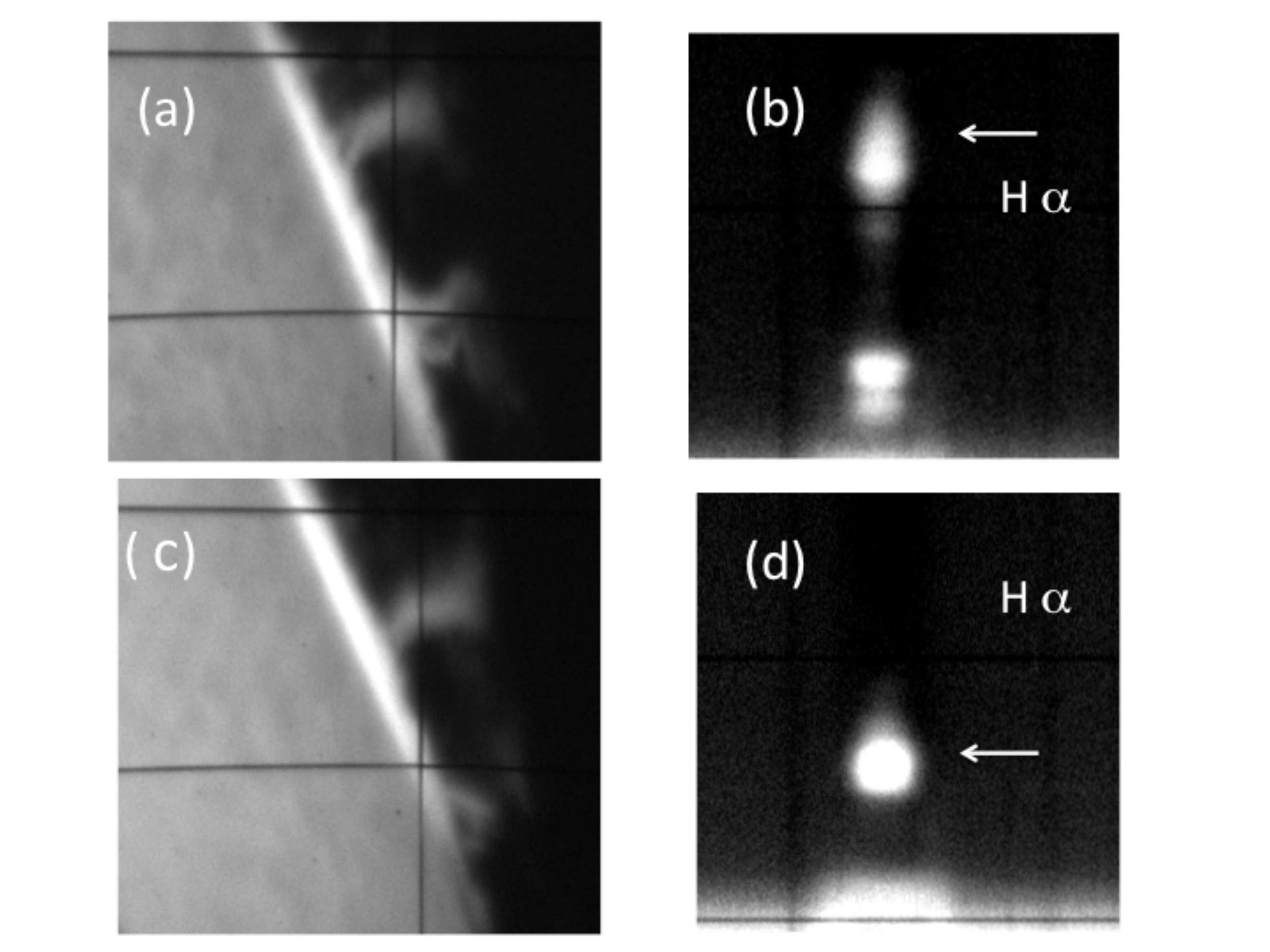}
      \caption{Tornado of 2014 May 23: H$\alpha$ observations of Ond\v{r}ejov observatory  at 13:00 UT -- 13:04 UT. \textit{(a)} and \textit{(c)} show the slit jaw images, \textit{(b)} and \textit{(d)} their respective spectra. The arrows indicate positions where the Doppler velocities have been measured (courtesy of P. Kotr\v{c}).  }
         \label{Ondrejov}
	\end{center}
   \end{figure}

We find the name ``tornado'' in the prominence classifications of \citet{Pettit1932}.  These tornado prominences  were described as such: ``The spiral form in a prominence sometimes gives it the appearance of a closely wound rope or screw''.  Recently,  dark prominence legs  observed in AIA 171 \AA\ images look as if they are rotating around a central axis \citep{Li2012}. In this paper we refer to these features as ``tornadoes'', regardless of if they are truly rotating or not. \textit{Hinode}/EIS revealed a split blueshift and redshift pattern with anti-symmetry along the vertical axis of the tornado{, which is perpendicular to the limb}, suggesting that there is rotation around this vertical axis \citep{Su2014,Levens2015}. 
{Obviously,} it is important to investigate the  magnetic field, which governs these tornadoes.
We had the opportunity to observe several such tornado-like prominences during the international observing campaigns in 2014 and 2015. These structures were mainly identified by their silhouette in the AIA 171 \AA\ and 193 \AA\ filters.  { We present the results of two tornado-like prominences which were also observed by THEMIS on May 23, 2014 and  July 15, 2014.  

{Context images of the tornado of May 23, 2014 are presented in Figures \ref{EIS} and \ref{Ondrejov}. The tornado of May 23, observed with \textit{Hinode}/EIS in \ion{Fe}{12} at 195 \AA\ presents blueshifts and redshifts ($\pm$ 3 km s$^{-1}$) as seen in  Figure \ref{EIS}. The characteristic tornado pattern is seen along the axis, even if it is not as clear as shown in the case of \cite{Su2014}. It is for this reason, along with the visual identification from AIA channels, that we refer to these features as `tornadoes'. 
In H$\alpha$, this prominence appears as two columns in the survey images of Meudon, Kanzelh\"{o}he and Ond\v{r}ejov. In high resolution \ion{Ca}{2} images obtained by \textit{Hinode}/SOT, horizontal strands or loops are observed from both sides of the columns (Figure \ref{EIS}).  
Ond\v{r}ejov observatory observed the prominence of May 23 2014 at 13:00 UT and 13:04 UT (see Figure \ref{Ondrejov}). T{he slit of the spectrograph, represented by a vertical line in the slit jaw,} crosses the prominence in two sections. The Doppler shifts along the slit  from low altitudes to higher altitudes are between 8.7 {km s$^{-1}$} and 7.1 {km s$^{-1}$}, which represents a difference of 1.6 {km s$^{-1}$},  in relative agreement with the measurements in \ion{He}{1} D$_3${, as can be seen in Figure \ref{themis_D3}}.

Figure \ref{fig6new} \textit{(a)}  presents the maps from the THEMIS observations on May 23 2014, obtained after inversion of the Stokes parameters recorded in the \ion{He}{1}  D$_3$  line (from top to bottom):  intensity, magnetic field strength, inclination, and azimuth.  The field strength is commonly below 15 G.  With a field strength upper limit reaching 50 G,  we see that values this high are achieved in {a few} isolated spots.
 
 As usual we focus on the inclination {with} respect to the local vertical. The brightest parts of the prominence have a mean inclination of 90$^\circ$ which means that the magnetic field  is mainly horizontal (Figure \ref{fig6new}).  Figure \ref{fig4new} \textit{(a)} shows the histogram which presents a main peak at 90$^\circ$ for the correctly inverted peaks (errors $<$ 10$^\circ$), this time with a more typical Gaussian shape, indicating that the departures from horizontal are random inversion errors rather than a true physical departure. The profles with errors $>$ 30$^\circ$ accumulate at two peaks on the sides, but contrary to the bubble case, this time their maxima are found at around 30$^\circ$ and 150$^\circ$. Since these inversions are not well defined, we should not hastily conclude on the presence of vertical fields, but rather on unusual magnetic topologies.

The tornado  of July 15, 2014 is well documented in \citet{Levens2016}.  Figure \ref{fig7new}  ({\it left column})   shows  the maps of July 15 2014 which have been obtained from THEMIS data. {We see, once again,}  two columns that are largely orange in the inclination maps, which suggests a horizontal direction for the magnetic field (90$^\circ$).
Looking in more detail, the histogram of the inclination (Figure \ref{fig4new} \textit{(b)}) for this observation again presents an intense and Gaussian peak centred around an inclination of 90$^\circ$, which corresponds to a horizontal magnetic field, but also shows  extended secondary peaks between 30$^\circ$ and 60$^\circ$ and between 120$^\circ$ and 150$^\circ$ for those cases with large inversion errors. 
The blue and white points in the inclination maps of Figure \ref{fig7new}  ({\it left column})  can now be seen to have error bars larger than 30$^\circ$ and, as before, we cannot conclude that there are vertical fields. After comparison with previous histograms of inclination, both in this work and in the cited literature, one is tempted to identify two distinct contributions in the histogram of inclinations with large errors. The first one is made of the now familiar peaks at 60$^\circ$ and 120$^\circ$, already found in the bubble case above, and analyzed by \citet{Schmieder2014} and \citet{Lopez2015}. These points can be interpreted as a background horizontal field mixed with a turbulent, isotropic, component. The second contribution would be made of two new peaks  centered at 30$^\circ$ and 150$^\circ$, as those found in Figure \ref{fig4new} \textit{(a)} (corresponding to the tornado observed on May 23, 2014). This solution, that we refer to as 30/150 here for commodity, appears to be associated to tornadoes in our datasets. Even when looking into larger datasets than those analyzed in this work, we only see it appearing associated to tornado structures and nowhere else. 
We conclude that  tornadoes present a singular and distinct magnetic topology that, when seen through the observation setup of THEMIS, results in profiles for which the PCA inversion code finds no appropriate solution.
Several cuts  (Figure \ref{fig6new} ({\it right column}), Figure \ref{fig7new} {(\it right column})) at selected heights through both tornadoes show the distribution of these three solutions over the prominence, as well as the actual values of the error bars. These errors can be seen to be as large as 60$^\circ$ when the 30/150 solution is retained, but in spite of this they do not appear to be due to the 90$^\circ$ ambiguity.
Still looking for the meaning of this distinct solution in tornadoes one can speculate that it may be due to the discrete nature of the database used for inversion and that, somehow, the Stokes profiles observed in tornadoes are simply missing from the database. While we cannot exclude this, the use of several different databases, all created with Monte Carlo techniques, reduce the probability of this explanation.

We advocate that the explanation to the 30/150 solution is similar to the one found for the peaks at 60$^\circ$ and 120$^\circ$. In that case it was found that a horizontal background field mixed with a turbulent isotropic field resulted in profiles that the PCA inversion code could not invert, but to which invariably it attributed solutions with either an inclination of 60$^\circ$ or 120$^\circ$, but with large error bars.  This is basically what we expect for the 30/150 solution:  a complex magnetic topology, mixing several magnetic fields in our resolution element (both temporal and spatial) and resulting in profiles which cannot be inverted by our model but for which the 30/150 solution is offered as the best match.  What such complex topology may be can only be suggested by 
the theoretical modelling of tornadoes.

\begin{figure*}
\begin{center}
\includegraphics[width=0.9\hsize,clip=true,trim=0 2cm 0 0]{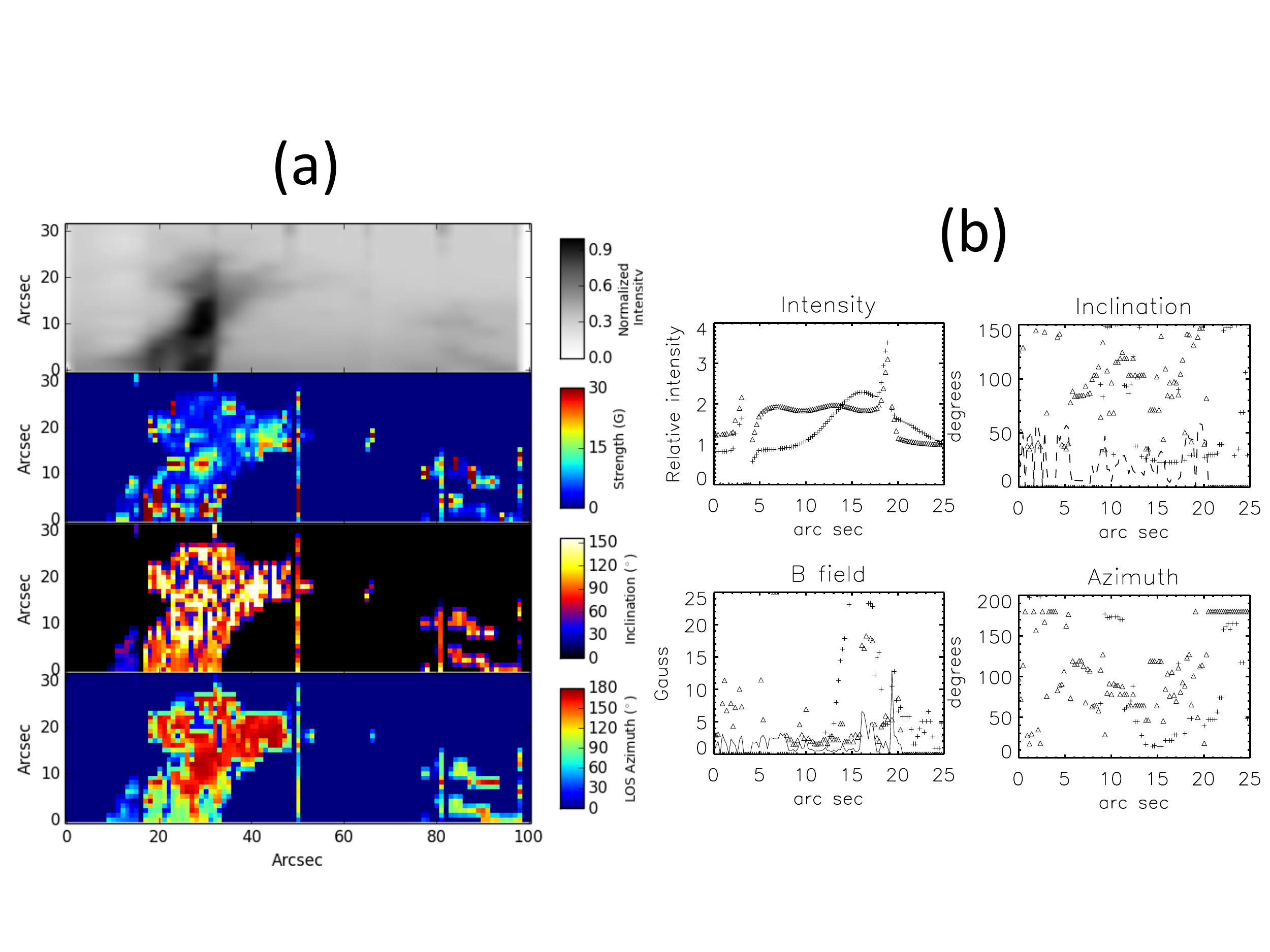}
\caption{THEMIS observation of the tornado of the 23 May 2014. \textit{(a)} From top to bottom: Intensity (reversed grey scale), magnetic field strength, inclination, and azimuth. \textit{(b)} Cuts through the prominence at two altitudes, showing again intensity (upper left), magnetic field strength (lower left), inclination (upper right), and azimuth (lower right). The two altitudes are at  12 \arcsec\ (plus signs) and 2 \arcsec\ (triangles). {The origin of the cuts in the left  panels is at x = 18\arcsec. The dashed lines in the panels of field strength and inclination represent the  computed errors.}}
\label{fig6new}
\end{center}
\end{figure*}

\begin{figure*}[t]
\begin{center}
\includegraphics[width=\hsize]{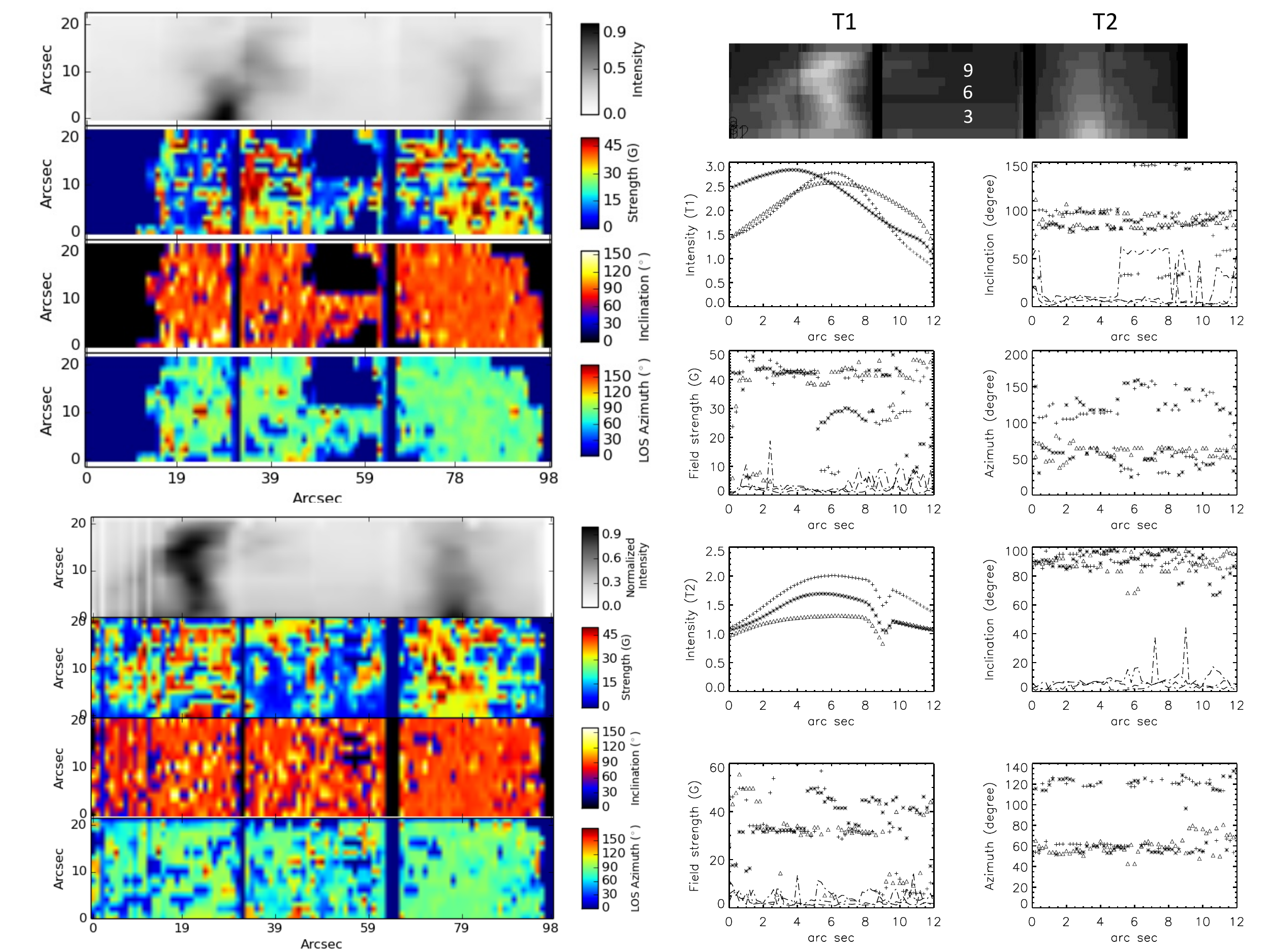}
\caption{Two tornadoes observed on 15 July 2014 by THEMIS. \textit{Left panels:} THEMIS maps from the two rasters, each showing the upper and lower parts of the tornado. Panels show, from top to bottom: Relative intensity (reversed grey scale), magnetic field strength, inclination, and azimuth. \textit{Right panels:} Cuts through the two tornadoes at three altitudes in the lower raster. Top panel shows positions of the cuts and identifies tornadoes as T1 (located at x = 14\arcsec\ -- 26\arcsec) and T2 (located at x = 70\arcsec\ -- 82\arcsec). Cuts are again relative intensity (upper left), magnetic field strength (lower left), inclination (upper right), and azimuth (lower right) for each. Cuts are taken at heights of 6\arcsec\ (cut 3, plus signs), 12\arcsec\ (cut 6, asterisks), and 18\arcsec\ (cut 9, triangles). {The origin of the cut through T1 is x = 14\arcsec, and that for T2 is at x = 70\arcsec. The solid and dashed lines in the panels of field strength and inclination  respectively represent the computed errors}.}
\label{fig7new}
\end{center}
\end{figure*}

\begin{figure*}
\begin{center}
\includegraphics[width=0.95\hsize]{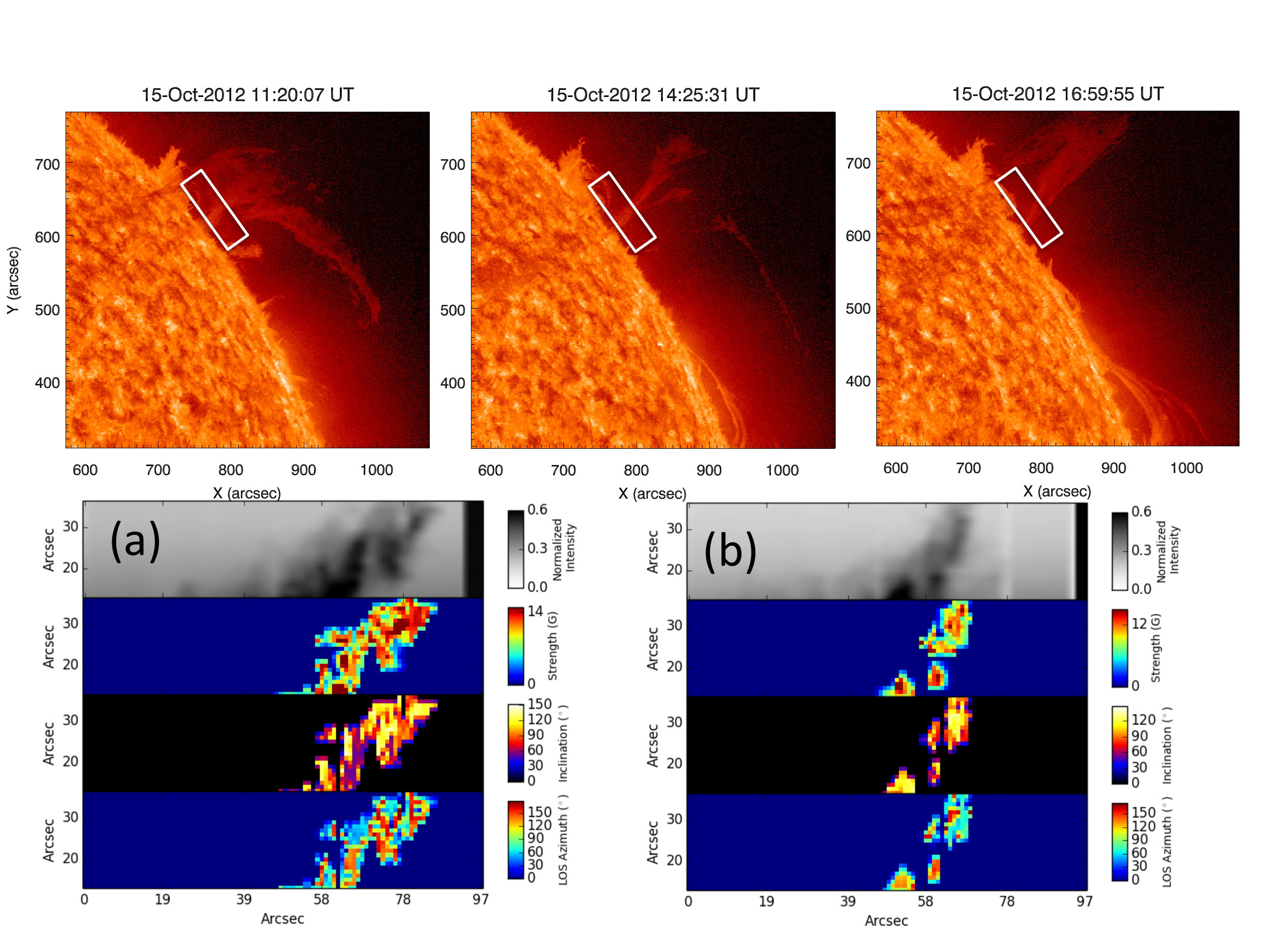}
\caption{Prominence eruption observed on 15 October 2012. \textit{Upper panels:} Images from AIA using the 304 \AA\ filter, showing the evolution of the prominence eruption from 11:21 UT to 16:56 UT. An animation is available online. White boxes indicate approximate FOV of the THEMIS rasters. \textit{Lower panels:} THEMIS observations of this eruption at \textit{(a)} 11:00 -- 12:00 UT, and \textit{(b)} 17:00 -- 18:00 UT. From top to bottom: Intensity (reversed grey scale), magnetic field strength, inclination, and azimuth.}
\label{fig8new}
\end{center}
\end{figure*}

\section{Magnetic field vector in an eruptive prominence}
{ 
The last case presented here is a prominence observed with THEMIS on 12 October, 2012 at PA = 311$^\circ$. The shape of the prominence was changing rapidly, and jet-like vertical structures were observed in the AIA  304 \AA\, images (Figure \ref{fig8new}, {\it top panels}). Two scans were made during this rapid evolution, the first one lasted from 11:00 UT -- 12:00 UT and covered 42\arcsec, from 8\arcsec\ off limb to 48\arcsec. The second scan was obtained 6 hours later, between 17:00 UT and 18:00 UT, and covered 26\arcsec. 
Figure \ref{fig8new} ({\it bottom panels}) shows the maps of intensity, magnetic field strength and orientation for the two scans. The rapid evolution seen in this prominence should cease any interpretation of these maps as static pictures of the prominence at a given time. The prominence was evolving as the scan proceeded, and the maps reflect this evolving picture. Nevertheless, our interest in this particular prominence is not to do with the spatial or temporal distribution of magnetic field properties, but rather in the fact that generally the inclination of the field
shows strong departures from horizontal. Figure \ref{fig4new} \textit{(d)}  shows the usual histogram for the inclination values and the 
difference with all other previous histograms is clear. Well inverted profiles, with no ambiguity in the inclination, are almost evenly distributed from 50$^\circ$ through 140$^\circ$, with two peaks around 50$^\circ$ and 130$^\circ$. We stress the fact that, contrary to the previous cases, these are  correct inversions, with error bars smaller than 10$^\circ$.  

Two conclusions can be extracted from this particular prominence. The first one is a confirmation of the conclusions from the  numerical tests on the inversion strategy: when vertical fields are present, we can measure them. If usually we only measure horizontal fields it is because only horizontal fields are present in the observed prominences. By vertical we mean non-horizontal, and unable to support plasma against gravity. Clearly a field inclined 50$^\circ$ from the vertical is not strictly vertical, but it is not a field that can support prominence plasma for a long time.

The second conclusion concerns the  magnetic field in an erupting prominence. Although a clear picture of the topology of this field cannot be retrieved from the data it is obvious that the prominence has abandoned the stable magnetic fields which are characteristic of quiescent prominences. In doing that we observe the magnetic field tilting more than 40$^\circ$  from the horizontal and the plasma erupting in jet-like structures. As expected, we confirm that in an erupting prominence the magnetic field has abandoned the horizontal geometry that allows it to support plasma.


\section{Discussions and conclusions}
{ 
We were able to observe nearly 200 prominences  during several observing campaigns from 2012 to 2015
 with the spectropolarimeter at the THEMIS telescope. For the majority of them, what we refer to as typical prominences, we
 detect  a horizontal  magnetic field  \citep{Lopez2015}.
 The inclination histograms of those typical objects  present a large maximum around 90$^\circ$, Gaussian in shape, with 
 a FWHM of roughly 10$^\circ$ corresponding to the inversion noise. That horizontal field is all that can be confidently measured in a typical prominence. The histograms also show secondary lobes around 60$^\circ$ and 120$^\circ$. Such inversions correspond to Stokes profiles of the \ion{He}{1} D$_3$ line for which no solution can be found in terms of a single vector magnetic field. Since those peculiar profiles are always very similar to one another, the inversion code always produces the same set of solutions as the nearest model it can find. We found that Stokes profiles consisting of the addition of a horizontal field plus a turbulent field in the same pixel can give a solution similar to those particular profiles \citep{Schmieder2014}.
 We cannot ensure that other complex models of the kind would not produce similar profiles, but since we observed chaotic plasma motions in the same places that those solutions were found lead us to remain with this interpretation: the two peaks at 60$^\circ$ and 120$^\circ$ in the inclination histograms correspond to turbulent fields on top of a background horizontal field.

{How can we trust the results from our inversion when faced with this problem of inhomogeneity? We can imagine two different scenarios: In the first, a prominence is made of the accumulation along the line of sight of optically thin structures that we can approximate as prominence threads. If the magnetic field is homogeneous enough along the line of sight, the average of the individual signals will not change the polarization patterns along the spectral line and our inversion will correctly fit the profiles. If, on the other hand, the magnetic field is very inhomogeneous, then the averaging will result in a ``nonsense'' profile in which the Hanle and Zeeman effects will average differently. The inversion code will not fit such profiles. 
In the present work we have identified such behaviour in our histograms of inclination and even proposed a turbulent scenario which reproduces one of the cases. A second scenario is that of an optically thick prominence. Each helium atom in the prominence is illuminated by two radiation fields: A first one made of the anisotropic cone of light from the photosphere that introduces  atomic polarization and which is what our inversion code computes, and a second one made of the isotropic illumination from neighbouring atoms in the prominence. 
This second radiation field introduces no atomic polarization. The conclusion here is therefore that the effect of the presence of radiative transfer is either a depolarization that does not affect the inferred magnetic field, either an averaging of signals over homogeneous enough structures to still be interpreted with our tools, or over inhomogeneous structures, in which case we obtain anomalous peaks in our histograms for which we try to propose sound models. 
Thus, we must also consider other strategies for interpreting our observations, such as detailed forward modelling of the observed scattering polarization signals using realistic 3D models of solar prominences and forward modeling techniques similar to those currently applied to the solar chromosphere \citep{Stepan2015}, taking into account the limited spatial and temporal resolution of the THEMIS observation.}
 
 Building upon this, we focused this paper on special cases of prominences to see whether their magnetic field presented {peculiar characteristics -- primarily those which presented a large error upon inversion}. We focused on three features of recent interest, namely bubbles, tornadoes, and  a jet-like eruptive prominence. 
 In all cases we found magnetic fields that, one way or another, departed from the general result on horizontal fields described above as corresponding to the typical prominence. Because of this we refer to these objects as atypical.

 The first feature studied are  ``bubbles'' observed below prominences. Two cases were identified and are presented here. The interest of the first one is that the same prominence has been studied and published in the literature before \citep{Shen2015}. Unfortunately, the signal-to-noise ratio is very low and {near to no conclusive results can be drawn from the data}. The only apparent result from this particular prominence with bubbles is that they appear to be surrounded by a relatively strong field (50G). The second prominence with bubbles we studied appears to confirm this result while also producing turbulent fields on top of a background horizontal field. This would confirm the prediction of the existence of highly magnetized plasma inside the  bubble, itself 
 outlined by a magnetic separatrix within the prominence  \citep{Dudik2012}. The rise of the bubble would be driven by this atypical magnetic field, and not by hot plasma inside the bubble \citep{Berger2014}.
 

The second feature that has been studied is tornadoes. Magnetically, all the cases present a common fact:  the magnetic field inclination presents a primarily horizontal direction, but has  two different kinds of secondary lobes in the histogram. One set of lobes is located between 60$^\circ$ and 120$^\circ$, and corresponds to the scenario described above -- a turbulent field on top of a background horizontal field. However, a second  set of lobes with maxima at  30$^\circ$ and 150$^\circ$ appears exclusively in the case of tornadoes. The field strength is generally around 15 G, but can  reach 40 -- 60 G in places. 

{The model presented by \citet{Luna2015} for the magnetic field in tornadoes is an interesting one, with a twisted magnetic field structure around a central axis. The magnetic field in the centre would be mostly vertical, with the field becoming more helical towards the outer edges. This model would provide unresolved mixtures of magnetic field orientations when run through the inversion code, however the resulting Stokes profiles would not be well interpreted by our present models, and would not result in the characteristic peaks that are seen in these observations. As is discussed in \citet{Levens2016} we also lack the mixed azimuth distribution that is implicit with the \citet{Luna2015} model.}


The final phenomena that was studied was a jet-like structure in  an eruptive  prominence.  We measure a non-horizontal  magnetic field in most of the pixels in this structure, with inclinations tilted to just 50$^\circ$ from the vertical. This structure confirms that whenever there are non-horizontal fields in prominences, we can measure them. Hence the typical horizontal field measured in typical prominences is not a bias of our measurement techniques but just the fact that horizontal fields are the dominant magnetic geometry in prominences. }

Atypical (i.e. bubbles, tornadoes, eruptive prominences) and typical prominences all are key signatures associated with highly stressed magnetic fields lying above photospheric polarity inversion lines. 
Understanding why and how it is that only some parts of these magnetic fields are associated with plasma emission remains one of the longest-standing debates in solar physics. 
Theoretical and numerical studies have given several constraints for and insights into the magnetic and plasma properties of such prominence materials. This has been done by looking at each independently, i.e. focusing either on the magnetic field or on radiative transfer modelling \citep[e.g.][]{DeVore2000,Aulanier2002,Karpen2005,Gunar2015}. 
\ \citet{Xia2014} and \citet{Terradas2015} were among the first to propose and study an MHD model of the formation of a self-consistent, plasma-carrying flux rope, with in situ condensation, forming a prominence. Observational studies, however, show that the magnetic field supporting prominence material is more complex than the relatively simple flux rope considered by \citeauthor{Xia2014} \citep[see e.g.][]{Aulanier1998,AulanierSch2002,vanBallegooijen2004,Jiang2014}. 
Similar MHD studies with more complex magnetic structures will open up new ways for investigating the formation of prominences. 

Combining multi-wavelength observations with high cadence and high spatial and spectral resolution is fundamental to diagnose properties of the plasma and magnetic fields in prominences. 
{Starting in early 2018, THEMIS will be upgraded with adaptive optics (AO) allowing it to improve  spectropolarimetric measurements.
 First  with an increase by  a factor of 2 in transmission  the maps  will be built in shorter times with the same polarimetric precision. Secondly the AO itself, which even beyond the limb will stabilise the image. It has been estimated that  the  spatial resolution will reach  0.25\arcsec  and at 1 arc minute from the correction centre of the AO, the image resolution  will be degraded  to 0.32\arcsec, which is a large improvement on the current 1\arcsec\ resolution.}
Meanwhile, the Coronal Multichannel Polarimeter \citep[CoMP;][]{Tomczyk2008} produces coronal spectropolarimetric measurements of forbidden lines in the near infrared. 
Such observations complement those of THEMIS by giving polarimetric data not only about the prominence, but also about its surroundings \citep{BakSteslicka2013,Rachmeler2013,Rachmeler2014}. This provides valuable information and constraints for magnetic field diagnostics (Dalmasse et al., \textit{in prep.}). 
Coordinated observations with instruments such as CoMP, IRIS, the upgraded THEMIS, ALMA, and later with DKIST (that will perform spectropolarimetric measurements such as the ones from THEMIS and CoMP) will provide unprecedented diagnostics to analyze the plasma and magnetic properties associated with prominences and bring stronger constraints for theoretical and numerical modelling.
 
\begin{acknowledgements}
We would like to thank the  team of THEMIS and S. Gun\'{a}r for acquiring the observations.
We thank  Xu  Zhi for providing the observations from the NVST of Kunming, and for coordinated with THEMIS. We thank P. Kotri\v{c} and Y. Kuprjakov for the observations in Ond\v{r}ejov.
Our thanks also go to  P. Mein, N. Mein, D. Crussaire and R. Lecocguen for the observations at the Meudon  Solar Tower with the MSDP instrument.
We would like to thank P. Heinzel for providing valuable feedback and helping improve this paper. 
We are grateful to the Hinode and IRIS planners who assisted with the observations in 2014 and 2015 as part of IHOP 255.
This  work was supported by the SOLARNET project (www.solarnet-east.eu), funded by the European Commission's FP7 Capacities Programme under the Grant Agreement 312495.
P.J.L. acknowledges support from an STFC Research Studentship ST/K502005/1. N.L. acknowledges support from STFC grant ST/L000741/1. K.D. acknowledges support from the Computational and Information Systems Laboratory and from the High Altitude Observatory. The National Center for Atmospheric Research is sponsored by the National Science Foundation.
 \textit{Hinode} is a Japanese mission developed and launched by ISAS/JAXA, with NAOJ as a domestic partnerand NASA and STFC (UK) as international partners. It is operated by these agencies in cooperation with the ESA and the NSC (Norway). 
 The AIA data are provided courtesy of NASA/\textit{SDO} and the AIA science team.

\end{acknowledgements}


\bibliographystyle{aa}
\bibliography{bibliography}


\end{document}